\def\gpath{./} 

  \documentclass[12pt]{article}
  \usepackage[T1]{fontenc} %
  \usepackage{pslatex} %

\usepackage{graphicx} %
\usepackage{natbib} %

\renewcommand\author{}
\newcommand\affiliation{}

\usepackage{graphicx}

\begin{document}


{\bf\large Coherent detection of orbital angular momentum in radio}
\bigskip



\author{L.~K.~S.\ Daldorff},
\affiliation{%
The Catholic University of America, Washington, DC, USA
}

\author{S.~M.\ Mohammadi},
\affiliation{
Infotrek, Stockholm, Sweden
}

\author{J.~E.~S.\ Bergman},
\affiliation{
Swedish Institute of Space Physics, Uppsala, Sweden
}

\author{B.\ Isham},
\affiliation{
Interamerican University of Puerto Rico, Bayam{\'o}n, Puerto Rico, USA
}

\author{M.~K.~T.\ Al-Nuaimi},
\affiliation{
Southeast University, Nanjing, China
}

\author{K.\ Forozesh},
\affiliation{
Upplysningscentralen AB, Stockholm, Sweden
}

\author{T.~D.\ Carozzi},
\affiliation{
Chalmers University of Technology, Gothenburg, Sweden
}




\bigskip

{\bf\em Abstract:} 
The angular momentum propagated by a beam of radiation has two contributions: spin angular momentum (SAM) and orbital angular momentum (OAM). SAM corresponds to wave polarisation, while OAM-carrying beams are characterized by a phase which is a function of azimuth. We demonstrate experimentally that radio beams propagating OAM can be generated and coherently detected using ordinary electric dipole antennas. The results presented here could pave the way for novel radio OAM applications in technology and science, including radio communication, passive remote sensing, and new types of active (continuous or pulsed transmission) electromagnetic measurements.

\bigskip

%

{\em Manuscript date: 23 December 2015} \smallskip


\bigskip
\hrule
\bigskip\bigskip










As early as 1909 the wave motions associated with a revolving shaft were studied by Poynting, who suggested, by analogy, that circularly-polarised light should carry angular momentum (OAM) 
\citep{Poynting:PRSL:1909}.
%
In 1936 Beth reported on an optical experiment \citep{Beth:PR:1936} in which he verified Poynting's prediction.
Since then others have studied electromagnetic OAM 
\citep{%
Carrara_1949_Nature,
diFrancia_1957_NuovCim,
Allen_1966_AmJPhys}, 
however 
it is only recently that electromagnetic OAM has 
been extensively studied and utilized, 
%
%
both theoretically 
\citep{%
Mair&al:N:2001,%
Thide&al:PRL:2007,
Mohammadi&al:IEEETAP:2010,
Mohammadi&al:RS:2010,
ThideA_2011_arXiv, 
BennisA_2013_EuCAP
} 
and experimentally, 
at 
visible
\citep{%
Allen&al:PRA:1992,
BeijersbergenA_1994_OpticsComm,
Paterson:PRL:2005,
LaveryA_2013_NJP,
LaveryA_2013_Science}, 
millimeter and microwave
\citep{%
Hajnal_1990_RSPSA, 
KristensenA_1994_OpticsComm, 
Turnball:OpticsComm:1996,
CourtialA_1998_PRL, 
JiangA_2009_CMC,
ChengA_2014_SciRep
}, %
and radio wavelengths
\citep{%
Leyser:PRL:2009, 
TamburiniA_2012_NJP
}. %
OAM has also been studied 
in electron beams 
\citep{McMorran:Science:2011}.  
%

%
%

In addition to energy and linear momentum, electromagnetic waves can propagate angular momentum 
to infinity. 
The total electromagnetic angular momentum has mode numbers $j = s + l$, where $s$ and $l$ are mode numbers, respectively, of the electromagnetic spin and orbital angular momentum\citep{Humblet:Physica:1943}. 
Classically, spin angular momentum manifests itself as wave polarisation, where mode numbers $s = \pm 1$ correspond to right- and left-hand circularly polarised modes
and $s = 0$ to linearly-polarised modes. 
Similarly, 
OAM-carrying beams are characterized by a phase which is a function of azimuth, 
and 
OAM mode numbers $l$ are classically manifested by 
a change in phase of 
$l \times 360^{\circ}$ 
around any arbitrary circle centred on the beam axis 
 \citep{Allen&al:PRA:1992,Turnball:OpticsComm:1996}.
%
%

We extend the radio angular momentum technique 
\citep{%
Carrara_1949_Nature, 
Allen_1966_AmJPhys, 
Hajnal_1990_RSPSA, 
KristensenA_1994_OpticsComm, 
CourtialA_1998_PRL, 
JiangA_2009_CMC, 
LiskaMeinke:NZ:1970, 
Leyser:PRL:2009, 
Mohammadi&al:IEEETAP:2010, Mohammadi&al:RS:2010, 
Thide&al:PRL:2007, 
ThideA_2011_arXiv
} %
by actively generating OAM radio beams having a variety of mode numbers using a simple antenna array 
\citep{Thide&al:PRL:2007}, 
and by using a phase-coherent technique
to detect and verify the transmitted modes 
\citep{Mohammadi&al:RS:2010}. 
%
%
The results presented here demonstrate
that radio beams propagating OAM can be generated
 \citep{Thide&al:PRL:2007,Mohammadi&al:IEEETAP:2010} 
and coherently detected
 \citep{Mohammadi&al:RS:2010} 
using ordinary electric dipole antennas.  
This 
could potentially pave the way for novel radio OAM applications in technology and science, including radio communication, radio and radar remote sensing, ionospheric radio diagnostics, and radio astronomy.
%
Note that, using current technology, it is only at radio frequencies that coherent measurements of electromagnetic fields, \emph{i.e.} measurement of both amplitude and phase, may be performed. 




An antenna array of $N$ identical sources at angular frequency $\omega$ and equal amplitudes has the array factor
\begin{equation}
\Psi=\sum_{n=1}^N \exp[-\mathrm{i}(\vec{k}\cdot\vec{x}_n-\phi_n)],
\end{equation}
where $\vec{k}$ is the wave vector, $\vec{x}_n$ is the position, $\phi_n$ is the phase of the $n$th emitter, and $\mathrm{i}$ represents $\sqrt{-1}$\citep{Mohammadi&al:IEEETAP:2010,Balanis:Book:2005,Chireix:1936,Knudsen:IRE:1953}. 
When the emitters are electric dipoles with dipole moments $\vec{d}$, the electromagnetic energy density $u=\varepsilon_0(|\vec{E}|^2+c^2|\vec{B}|^2)/2$, where $\vec{E}$ and $\vec{B}$ are the electric and magnetic fields, becomes
\begin{equation}
u=\frac{k^2 |\vec{k}\times\vec{d}|^2 |\Psi|^2}{\epsilon_0 (4\pi r)^2}+O(r^{-3})\,,
\end{equation}
where $\varepsilon_0$ is the vacuum permittivity and $r=|\vec{x}|$ is the radial distance from the centre of the array, where $\vec{x}$ is the position vector.
The total angular momentum density is defined as $\vec{h}=\vec{x}\times\vec{g}$, where $\vec{g}=\varepsilon_0\mathrm{Re}\{\vec{E}\times\vec{B}^{\ast}\}$ is the linear momentum density. In our case we have linearly-polarised dipoles, 
\emph{i.e.} $\vec{d}\times\vec{d}^{\ast}=\vec{0}$ 
and $s = 0$, 
so that the total angular momentum $j = l + s = l$.
Hence, 
only the OAM contributes to the total 
angular momentum 
density, 
\begin{equation}
\vec{h}=\frac{u}{\omega}\frac{\Re{\Psi^\ast\widehat{\vec{L}}\Psi}}{|\Psi|^2}+\vec{O}(r^{-3})\,,\label{eq:h}
\end{equation}
where $\widehat{\vec{L}}=-\mathrm{i}(\vec{x}\times\nabla)$ is the OAM operator.
Since the OAM operator in Eq.~\ref{eq:h} only operates on the array factor $\Psi$, the only contribution to the angular momentum of the fields will be from the phasing of the elements and the geometry of the array, not the individual antenna elements, so long as the elements are much smaller than the size of the array.

We have used a circular array of $N=6$ elements placed in the $xy$ plane with emitters distributed equidistantly around the perimeter of a circle of radius $a$ as shown in 
fig.~1,
and phased such that $\phi_n = 2\pi ln/N$, with $l$ an integer. The array factor becomes
\begin{equation}
\Psi_l= N (-\mathrm{i})^l \exp(\mathrm{i}l\varphi) J_l(ka\sin\theta),\label{eq:FFT}
\end{equation}
 where, $\theta$ and $\varphi$ are the spherical polar and azimuthal angles, respectively, and $J_l$ is the Bessel function. The array factor
$\Psi_l$
 contains the phasor $\exp(\mathrm{i}l\varphi)$ which is a characteristic of OAM beams\citep{Allen&al:PRA:1992}. The phase factor varies around the beam axis and the OAM number, $l$, corresponds to a Fourier component. Thus, by measuring the phase of a single field component, the OAM modes can be separated by a spatial Fourier transform about the $z$ axis \citep{Elias:AA:2008}.

\begin{figure}[tp] \centering \includegraphics[scale=0.14]{\gpath/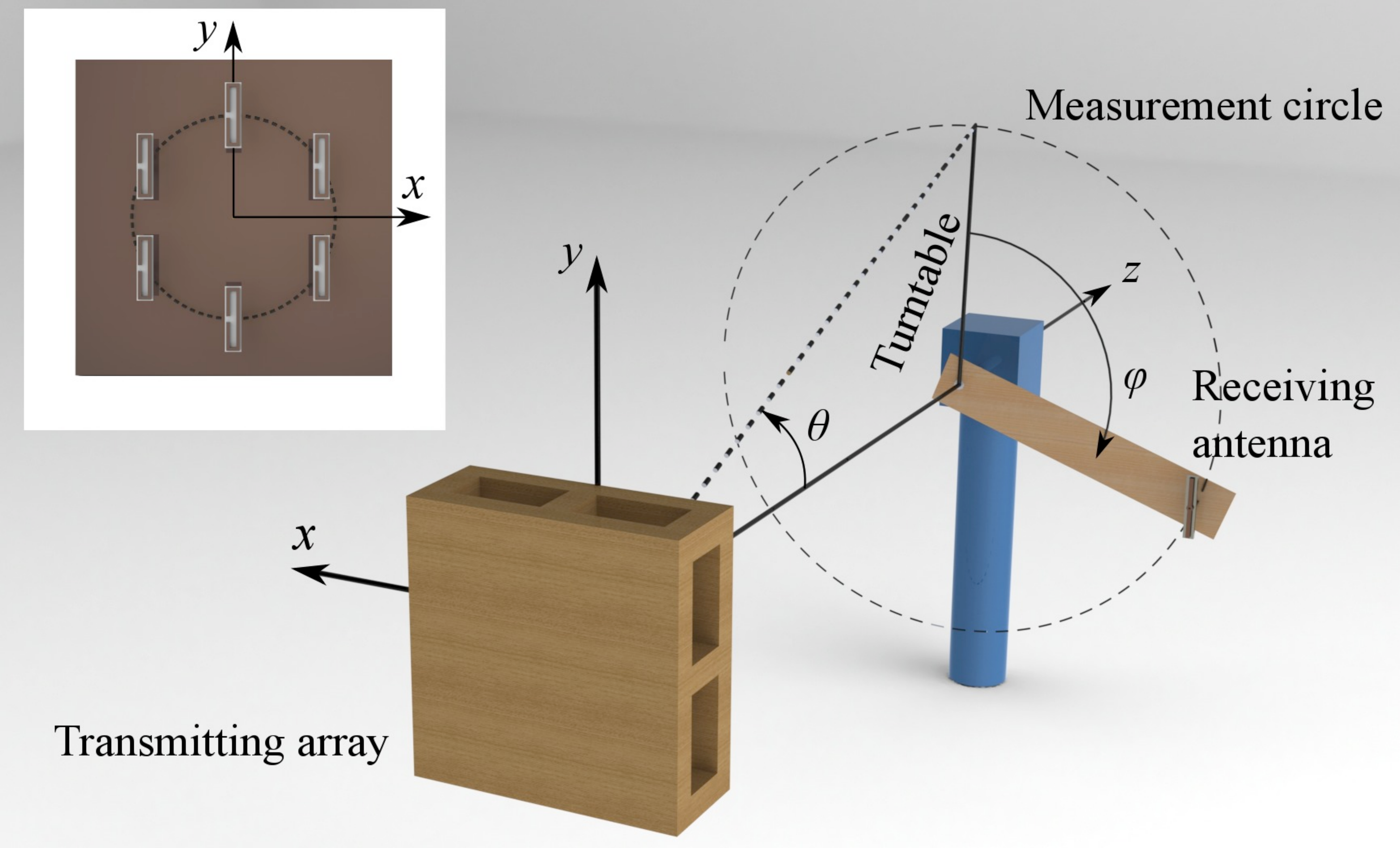} 
\caption{
\label{fig:1}
The experimental setup. 
The six-element transmitting antenna array is at the lower left, at the origin of the coordinate system, with a front view of the array shown in the inset at top left.
The receiving antenna is located to the right, at the end of a rotating arm.
The arm was connected to the turntable,
the pivot of which was located on the $z$ axis,
and which
was supported by a vertical pillar (blue) connected to the floor.
The transmitting antennas were half-wavelength folded dipoles directed along the vertical, parallel to the $y$ axis, positioned equidistantly around a circle of radius $a = 0.75 \, \lambda$
(where $\lambda$ is the wavelength) and placed parallel to,
and at a distance of
$0.013$ m ($0.1 \, \lambda$)
in front of,
a conducting plane.
The $xy$ plane, and the dashed circle of the transmitting array
(see inset),
were parallel to a chamber wall, and the $z$ axis was perpendicular to the same wall.
By feeding the transmitting antennas with equal amplitudes and with a phase difference between consecutive elements
of $0^\circ$, $\pm60^\circ$, and $\pm 120^\circ$, OAM modes $l = 0$, $l = \pm1$, and $l = \pm2$, respectively, were generated.
The centre axis of the horizontally-transmitted beam was aligned with the rotation axis of the
rotating arm.
The receiving antenna was attached to the arm, so that it rotated around the beam axis at a radius of 1.00 meter, thus avoiding the null at the centre of the
OAM 
beams. 
The measurement plane was located 4.82 m ($38.3 \, \lambda$) from the transmitting array;
the angle $\theta$ was $11.7^\circ$.
During the measurements, the arm was rotated in a vertical plane, parallel to the $xy$ plane.
The receiving antenna was vertically-aligned and measured the $y$ component of the transmitted electric field.
} \end{figure}


An array of $N=6$ folded half-wavelength dipole antennas was constructed,
as shown in the inset in 
fig.~1.
%
The transmitting antennas were numerically simulated using the method-of-moments-based EM simulator Ansoft Designer SV 
and 
fabricated using LPKF prototyping PCB milling machine 
on RO4003C dielectric material, which has dielectric constant 3.38, dissipation factor $\tan(\delta)=0.0027$, and copper thickness 0.018 mm \citep{RogersCorporation_2015_datasheet}. 
The antennas were fed with equal amplitudes and with phase shifts of $\delta\phi=0^{\circ}$, $\pm 60^{\circ}$, and $\pm 120^{\circ}$ between consecutive elements, such that the array generated beams carrying OAM modes $l=0$, $\pm 1$, and $\pm 2$, respectively.
The radiation patterns of the simulated beams are shown in 
fig.~2.
In the simulations, ordinary dipoles were used instead of folded dipoles for simplicity.
%
NEC-4 \citep{Adler_1993_ACESN} was used for modeling the beams. 
\begin{figure*}[tp] \centering \includegraphics[scale=0.45]{\gpath/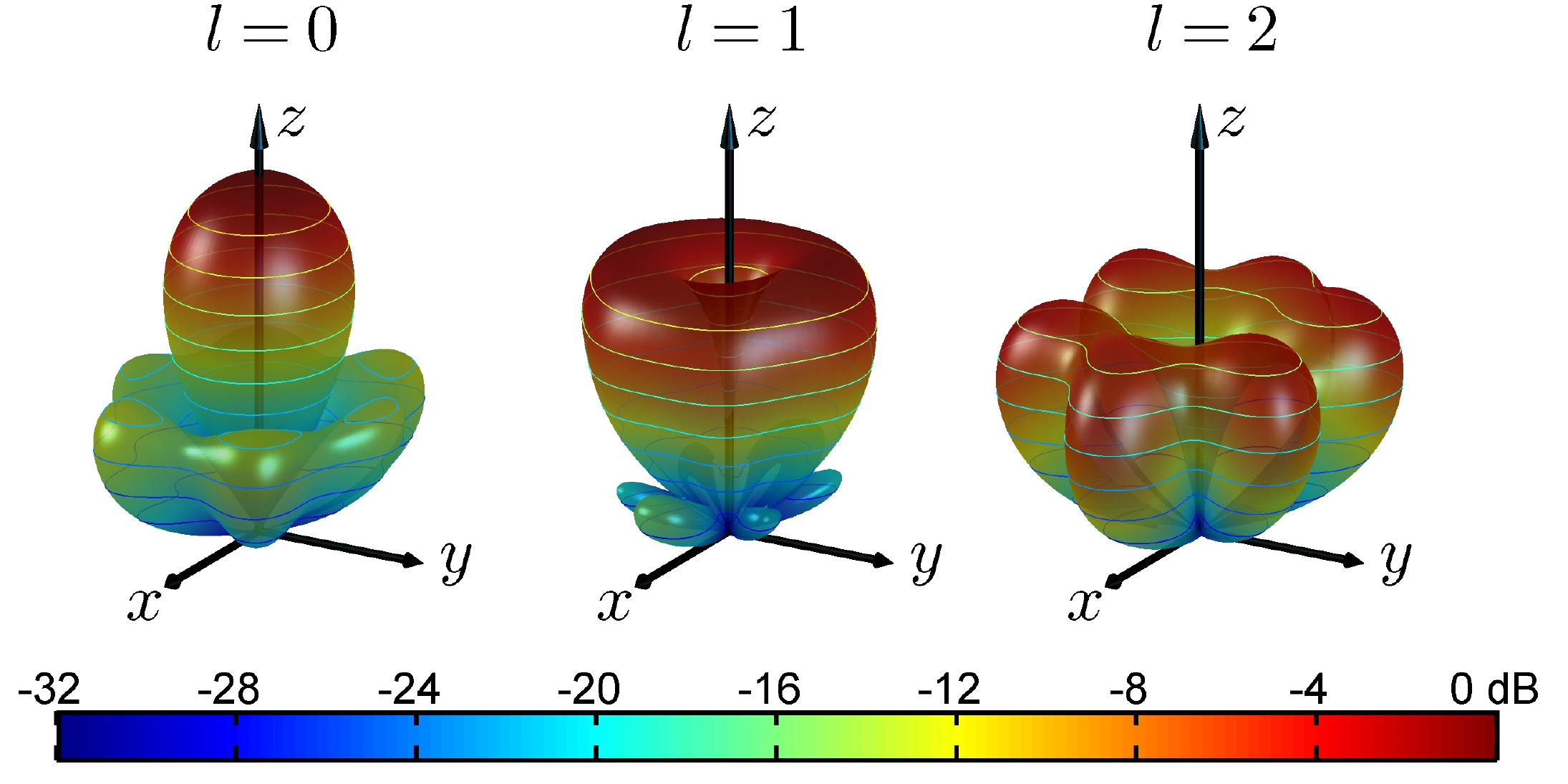}
\caption{
\label{fig:2} 
Simulated radiation patterns 
for the $l=0$, $\pm1$, and $\pm2$ radio beams. 
For $l \ne 0$ 
the beams exhibit the characteristic null along the beam axis ($\theta=0^\circ$)\citep{Thide&al:PRL:2007, Mohammadi&al:IEEETAP:2010, Mohammadi&al:RS:2010,Josefsson&Persson:2006}.
Simulations, analytical calculations, and measurements show that the 
$l=0$ and $\pm1$ 
beams have maxima at $\theta=0^{\circ}$ and $\theta=22^{\circ}$, respectively.
For $l = \pm2$, 
the angle of the beam
varied significantly around the beam axis. This is because 
$l = \pm2$ 
is much closer to the theoretical limit in the inequality $|l|<N/2$ for unambiguous transmission of OAM modes; the OAM beam will degrade as the limit is approached.
Note also that the amplitude patterns for 
$l=\pm2$ 
exhibit ripples.
Small ripples were also present for $l=\pm1$ but are not visible in the figure. 
Ripples generally appear when $|l|$ approaches the upper limit of $N/2$.
The color scale and contour lines indicate power density for each beam relative to the maximum of the $l = 0$ beam; the interval between contour lines is 3.2 dB.
} \end{figure*} 

As can be seen in the phase plots
in the upper panels of 
fig.~3,
when looking up the beam towards the transmitting array 
(\emph{i.e.}
towards the page in 
fig.~3), 
the lines of constant phase spiral out from the beam axis counterclockwise for $l>0$ (right-handed) and clockwise (left-handed) for $l<0$.
One complete turn at constant radius around the beam axis
changes the phase of the field components by $l\times 360^{\circ}$.
Accordingly, for the measurements presented here, the phase of $E_y$, which is the strongest electric field component, was measured at a constant radius from the beam axis \citep{Mohammadi&al:RS:2010}.

\begin{figure*}[tp] \centering \includegraphics[scale=0.35]{\gpath/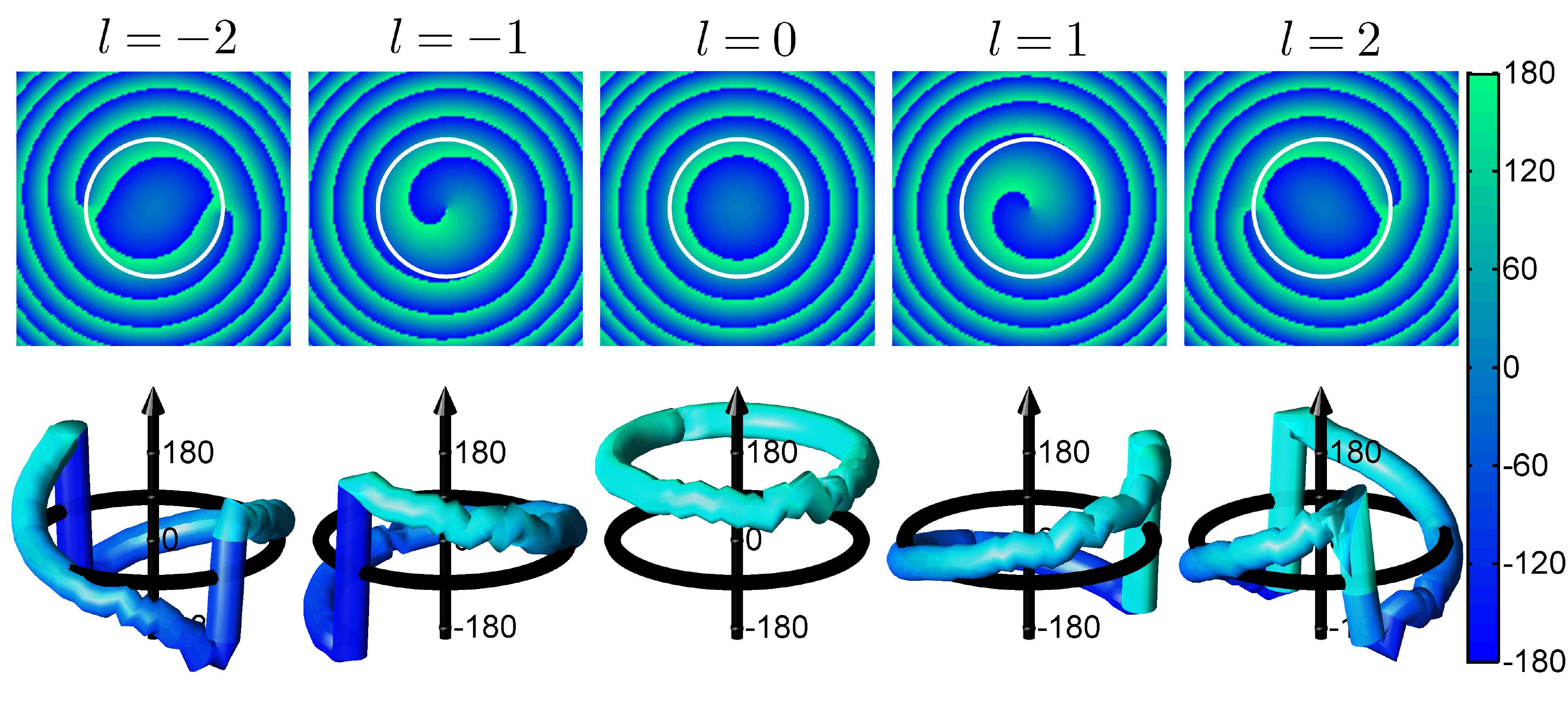} 
\caption{
\label{fig:3}  
Simulated and measured phase distributions in the OAM beam. 
In the upper panels the positive $z$ axis is out of the paper, in the lower panels the positive $z$ axis is oriented upwards.
The upper panels show the simulated OAM beam phase distribution for the $y$ component of the electric field at the measurement plane.
From left to right, the panels show the phase for $l = -2$, $-1$, $0$, $1$, and $2$, as seen looking along the beam axis toward the transmitting array, \emph{i.e.} in the negative $z$ direction.
The white circle indicates the position of the measurement circle.
The lower panels show the measured phase of the $y$ component of the electric field. The measured phase is indicated by both colour and vertical displacement. The OAM mode is found by counting the number of branch cuts from $-180^\circ$ to $+180^\circ$, \emph{i.e.}  for $|l| = 0$, $1$, and $2$ we have $0$, $1$, and $2$ branch cuts, respectively. The orientation of the phase slope gives the sign of the measured OAM mode. Declining phase values when the $xy$ plane is traversed in a right-handed sense indicate a negative
mode, and increasing values
a positive mode. 
The distortions visible in the measured phase are due to increased reflection 
when the receiving antenna is close to the floor. 
The color bar at right shows phase in degrees; both the upper and lower panels use the same color scale.
} \end{figure*}

The measurements took place in an anechoic chamber at a frequency of $2.383$ GHz (corresponding to 
a wavelength, $\lambda$, of  $0.126$ m) 
and at a distance of $z = 4.82$ m ($38.3\ \lambda$) from the transmitting array and hence well into the far-field region. 
The $E_y$ field was measured with a half-wave dipole antenna made from brass bars, connected to a $\lambda/4$ bazooka balun fabricated from rigid co-axial cable, and placed at the tip of a one-meter-long rotating arm, which was mounted on a pillar attached to the floor, see fig.~1. 
%
The amplitude and phase of $E_y$ were measured every $4^{\circ}$, for $l=0$, $\pm 1$, and $\pm 2$.
Repeated measurements for $l=1$ showed that the beam amplitude and phase were stable.

For $l = 0$ the measured data show a sinusoidally-oscillating, rather than constant, phase around the measurement circle.  
This oscillation was found to be produced by a 
small
misalignment between the centre of the transmitted beam and the centre of the measurement circle.  
This misalignment also produced a periodic phase oscillation in the $l = \pm1$ and $\pm2$ data.  The $l = 0$ measurement was used to correct the phase data for all modes.
The ripples in the phase plots in the lower panels of 
fig.~3
indicate the effect of spatial reflections of the signal and that the transmission was not a pure OAM mode,  as can be expected from the variation of the beam maximum around the beam axis in the radiation pattern for $l=1$ and $2$ in 
fig.~2.

Fig.~4 
shows the
measured $l$ spectra, computed by means of a spatial Fourier transform about the $z$ axis (see Eq.~\ref{eq:FFT} above).
Only minor errors arise at the transmitting side, where the feeding network delivered a maximum phase error of $3^{\circ}$ and a maximum amplitude difference of $0.03$\,dB to the six transmitting antennas.  These antennas were well-tuned and less than $2\%$ of the power was reflected 
back to the transmitter.
The spread in the spectrum is primarily due to the finite number of transmit antennas and the reflections in the measurement chamber.
The spectra
confirm that the intended OAM modes were transmitted and correctly detected.

\begin{figure}[tp] \centering \includegraphics[scale=0.5]{\gpath/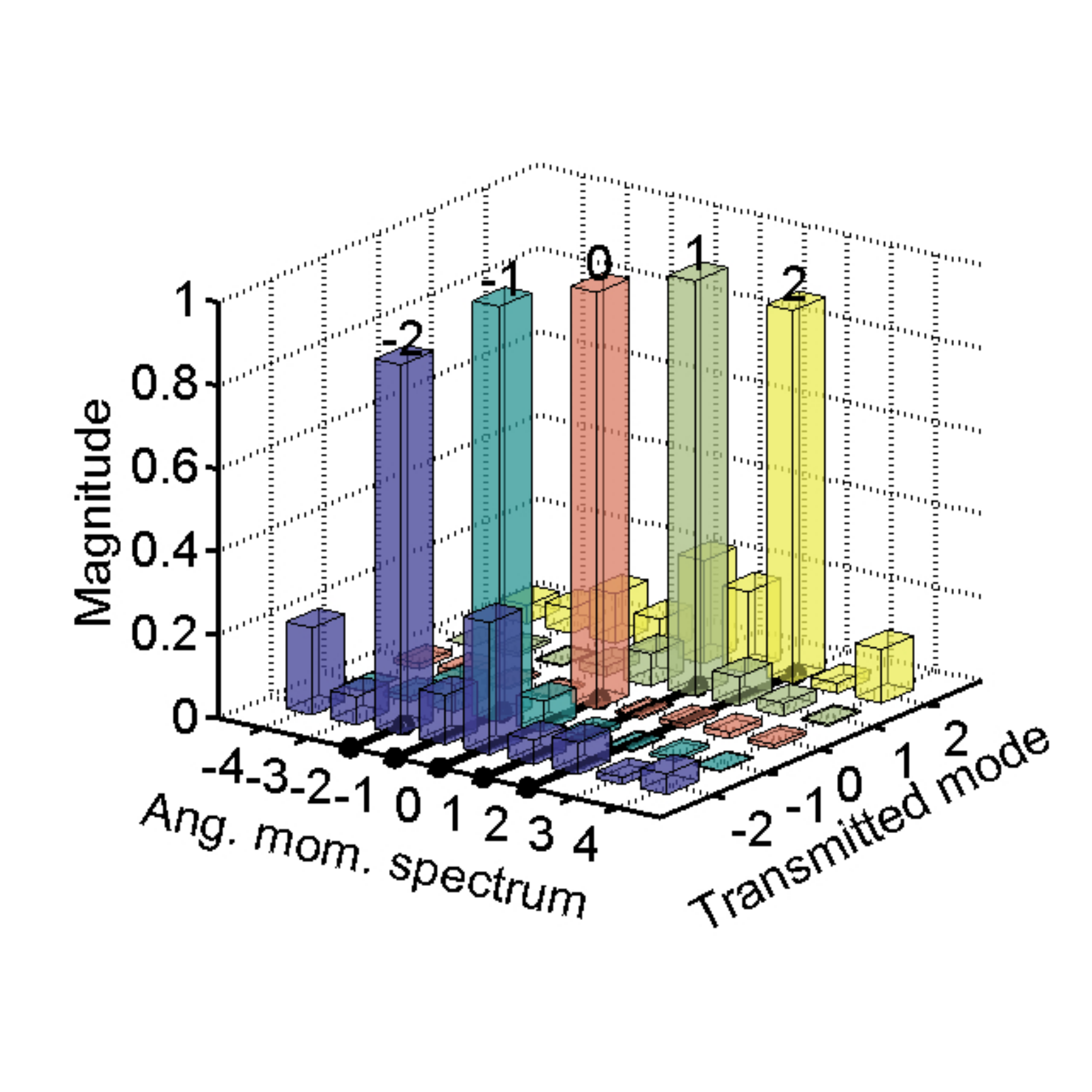}
\caption{ 
\label{fig:4}  
Angular momentum spectra. 
Five data sets of 90 phase measurements each were taken every 4$^{\circ}$ around the measurement circle.
Each set was Fourier transformed, from $\varphi$ space to $l$ space. The magnitudes of the Fourier components corresponding to OAM mode numbers $-4\le l \le 4$ are shown. The spectra show some leakage between mode numbers but the peaks are well correlated with the
intended
OAM modes, $-2\le l \le 2$. The magnitude scale is normalized for each mode such that the sum of the magnitudes from $l = -45$ to $l = +44$ is equal to $1$.
} \end{figure}


In summary, we have generated several radio OAM modes using a circular antenna array and successfully verified these modes via measurements of a single electric field component using an ordinary electric half-wave dipole; 
the detection of OAM does not require the use of a receiving array or a measurement of the full electric and magnetic field vectors.  
In addition, we have demonstrated that it is sufficient to measure the phase rotation to determine the OAM mode of a radio beam. 
Although the power pattern for the $|l| = 2$ beams was not ideal, the spatial phase patterns are insensitive to imperfections in the beam, 
and the measured phase depends only weakly on 
azimuthal 
field strength fluctuations
and on the accuracy of the circular measurement path around the beam axis \citep{Mohammadi&al:RS:2010, Mohammadi&al:IEEETAP:2010}.

%


This experiment opens up new possibilities for radio communications and for radio and radar remote sensing of rotational phenomena in the atmosphere and space.
Backscatter from OAM-modulated radar observations of the ionosphere can be analyzed using the methods presented here to diagnose possible rotational properties of plasma processes.
Artificially-induced and natural radio emissions from the ionospheric plasma \citep{Leyser:SSR:2001,LaBelleTreumann:SSR:2002,Leyser:PRL:2009} could be analyzed for possible OAM effects, lending clues to plasma processes.
The 
large-array radio telescopes 
recently constructed and being planned
could be used to search for signs of electromagnetic OAM arriving from astrophysical plasmas \citep{Harwit_2003_ApJ,Elias:AA:2008}.
%
Each OAM mode can act as an independent channel for transmission and reception, suggesting the possibility of increasing the information transfer rate within existing measurement and communications bands 
\citep{EdforsJohansson_2012_TAP,
TamburiniA_2012_NJP}. 
The information carried in OAM modes may also 
allowing radio and radar imaging at resolutions below the Rayleigh limit
\citep{LianlinFang_2013_PRE}.
In all applications, the reception of weak radio signals located close to an undesired strong source might be improved by orienting the the central null in an OAM receiving antenna towards the undesired source.
For example, radio observations of the solar corona could be performed by placing the central null in an OAM beam over the radio-bright disk of the sun.

{\em Acknowledgments:}
We thank Roger Karlsson, 
Lennart {\AA}hl{\'e}n, Walter Puccio, Sven-Erik
Jansson, Farid Shiva, Thomas Oswald, Erland Cassel, Shi Cheng, and Johan Lindberg for
technical support and advice.  
We thank Anders Rydberg and Anders Ahl\'en 
for loaning equipment
and for allowing access to the new {\AA}ngstr{\"o}m Laboratory antenna chamber, 
which was funded by the Knut and Alice Wallenberg Foundation. 
We thank the electronics lab of the Swedish Institute of Space Physics in Uppsala for 
granting access to their facilities, 
lending equipment, 
and for covering 
the electrical equipment expenses. 
We thank Masih Noor at the Center for Accelerator and Instrument Development 
for help with the CAD drawing for Figure \ref{fig:1} 
and Anders Hast and Martin Ericsson at UPPMAX for valuable help and feedback in visualizing the antenna patterns. 
We thank Bo Thid{\'e} for providing the inspiration for this work. 
The construction materials and personnel hours required for the experiment were privately financed by the authors.  
S.\ M.\ M.\ thanks Interamerican University of Puerto Rico for its hospitality during part of this work.
B.\ I.\ was supported by ARO grant 
W911NF-10-1-0002. 
We also thank the referees for constructive and helpful comments. 

\bibliographystyle{agufull04tg}

\begin{thebibliography}{39}
\providecommand{\natexlab}[1]{#1}
\expandafter\ifx\csname urlstyle\endcsname\relax
  \providecommand{\doi}[2]{{#1}:\discretionary{}{}{}#2}\else
  \providecommand{\doi}[1]{{#1}:\discretionary{}{}{}\begingroup
  \urlstyle{rm}\Url}\fi

\bibitem[{\textit{Adler}(1993)}]{Adler_1993_ACESN}
Adler, D. (1993), Information on the history and availability of {NEC-MOM}
  codes for {PC}s and unix, \textit{Applied Computational Electromagnetics
  Society Newsletter}, \textit{8}(3), 8--10,
  http://\-www.\-nec2.\-org/\-nec\_\-hist.\-txt, accessed 2015-08-30.

\bibitem[{\textit{Allen et~al.}(1992)\textit{Allen, Beijersbergen, Spreeuw, and
  Woerdman}}]{Allen&al:PRA:1992}
Allen, L., M.~W. Beijersbergen, R.~J.~C. Spreeuw, and J.~P. Woerdman (1992),
  Optical angular momentum of light and the transformation of
  {L}aguerre-{G}auss laser modes, \textit{Phys. Rev. A Gen. Phys.},
  \textit{45}, 8185--8189.

\bibitem[{\textit{{Allen}}(1966)}]{Allen_1966_AmJPhys}
{Allen}, P.~J. (1966), A radiation torque experiment, \textit{Am. J. Phys.},
  \textit{34}, 1185--1192, \doi{doi}{http://dx.doi.org/10.1119/1.1972585}.

\bibitem[{\textit{Balanis}(2005)}]{Balanis:Book:2005}
Balanis, C.~A. (2005), \textit{Antenna Theory}, 3 ed., John Wiley and Sons, New
  York.

\bibitem[{\textit{{Beijersbergen} et~al.}(1994)\textit{{Beijersbergen},
  {Coerwinkel}, {Kristensen}, and {Woerdman}}}]{BeijersbergenA_1994_OpticsComm}
{Beijersbergen}, M.~W., R.~P.~C. {Coerwinkel}, M.~{Kristensen}, and J.~P.
  {Woerdman} (1994), Helical-wavefront laser beams produced with a spiral
  phaseplate, \textit{Optics Comm.}, \textit{112}, 321--327,
  \doi{doi}{http://dx.doi.org/10.1016/0030-4018(94)90638-6}.

\bibitem[{\textit{Bennis et~al.}(2013)\textit{Bennis, Niemiec, Brousseau,
  Mahdjoubi, and Emile}}]{BennisA_2013_EuCAP}
Bennis, A., R.~Niemiec, C.~Brousseau, K.~Mahdjoubi, and O.~Emile (2013), Flat
  plate for oam generation in the millimeter band, in \textit{7th European
  Conference on Antennas and Propagation (EuCAP)}, art.\ no.\ 6546903.

\bibitem[{\textit{Beth}(1936)}]{Beth:PR:1936}
Beth, R.~A. (1936), Mechanical detection and measurement of the angular
  momentum of light, \textit{Phys. Rev.}, \textit{50}(2), 115--125.

\bibitem[{\textit{{Carrara}}(1949)}]{Carrara_1949_Nature}
{Carrara}, N. (1949), Torque and angular momentum of centimetre electromagnetic
  waves, \textit{Nature}, \textit{164}, 882--884,
  \doi{doi}{http://dx.doi.org/10.1038/164882c0}.

\bibitem[{\textit{Cheng et~al.}(2014)\textit{Cheng, Hong, and
  Hao}}]{ChengA_2014_SciRep}
Cheng, L., W.~Hong, and Z.-C. Hao (2014), Generation of electromagnetic waves
  with arbitrary orbital angular momentum modes, \textit{Sci. Rep.},
  \textit{4}(1484), 5 pp., \doi{doi}{http://dx.doi.org/10.1038/srep04814}.

\bibitem[{\textit{Chireix}(1936)}]{Chireix:1936}
Chireix, H. (1936), {A}ntennas \'a rayonnement z\'enital r\'educit,
  \textit{L'Onde \'Elec.}, \textit{15}, 440--456.

\bibitem[{\textit{{Courtial} et~al.}(1998)\textit{{Courtial}, {Dholakia},
  {Robertson}, {Allen}, and {Padgett}}}]{CourtialA_1998_PRL}
{Courtial}, J., K.~{Dholakia}, D.~A. {Robertson}, L.~{Allen}, and M.~J.
  {Padgett} (1998), Measurement of the rotational frequency shift imparted to a
  rotating light beam possessing orbital angular momentum, \textit{Phys. Rev.
  Lett.}, \textit{80}, 3217--3219,
  \doi{doi}{http://dx.doi.org/10.1103/PhysRevLett.80.3217}.

\bibitem[{\textit{{di Francia}}(1957)}]{diFrancia_1957_NuovCim}
{di Francia}, G.~T. (1957), On a macroscopic measurement of the spin of
  electromagnetic radiation, \textit{Nuov. Cim.}, \textit{6}, 150--167.

\bibitem[{\textit{{Edfors} and {Johansson}}(2012)}]{EdforsJohansson_2012_TAP}
{Edfors}, O., and A.~J. {Johansson} (2012), Is orbital angular momentum
  ({OAM})-based radio communication an unexploited area?, \textit{IEEE
  Transactions on Antennas and Propagation}, \textit{60}, 1126--1131,
  \doi{doi}{http://dx.doi.org/10.1109/TAP.2011.2173142}.

\bibitem[{\textit{{E}{l}{i}{a}{s}{\ }{I}{I}}(2008)}]{Elias:AA:2008}
{E}{l}{i}{a}{s}{\ }{I}{I}, N.~M. (2008), Photon orbital angular momentum in
  astronomy, \textit{Astron. Astrophys.}, \textit{492}(3), 883--922,
  \doi{doi}{http://dx.doi.org/10.1051/0004-6361:200809791}.

\bibitem[{\textit{{Hajnal}}(1990)}]{Hajnal_1990_RSPSA}
{Hajnal}, J.~V. (1990), Observations of singularities in the electric and
  magnetic fields of freely propagating microwaves, \textit{Proc. Roy. Soc.
  London A}, \textit{430}, 413--421,
  \doi{doi}{http://dx.doi.org/10.1098/rspa.1990.0097}.

\bibitem[{\textit{{Harwit}}(2003)}]{Harwit_2003_ApJ}
{Harwit}, M. (2003), Photon orbital angular momentum in astrophysics,
  \textit{Astrophys. J.}, \textit{597}, 1266--1270,
  \doi{doi}{http://dx.doi.org/10.1086/378623}.

\bibitem[{\textit{Humblet}(1943)}]{Humblet:Physica:1943}
Humblet, J. (1943), Sur le moment d'impulsion d'une onde
  {\'e}lectromagn{\'e}tique, \textit{Physica}, \textit{X}(7), 585--603.

\bibitem[{\textit{Jiang et~al.}(2009)\textit{Jiang, He, and
  Li}}]{JiangA_2009_CMC}
Jiang, Y., Y.~He, and F.~Li (2009), Wireless communications using
  millimeter-wave beams carrying orbital angular momentum, in
  \textit{Conference on Communications and Mobile Computing}, vol.~1, pp.
  495--500, Kunming, Yunnan, China.

\bibitem[{\textit{Josefsson and Persson}(2006)}]{Josefsson&Persson:2006}
Josefsson, L., and P.~Persson (2006), \textit{{C}onformal {A}rray {A}ntenna
  {T}heory and {D}esign}, Wiley-Interscience.

\bibitem[{\textit{Knudsen}(1953)}]{Knudsen:IRE:1953}
Knudsen, H.~L. (1953), {T}he field radiated by a ring quasi-array of an
  infinite number of tangential or radial dipoles, \textit{Proc. IRE},
  \textit{41}, 781--789.

\bibitem[{\textit{{Kristensen} et~al.}(1994)\textit{{Kristensen},
  {Beijersbergen}, and {Woerdman}}}]{KristensenA_1994_OpticsComm}
{Kristensen}, M., M.~W. {Beijersbergen}, and J.~P. {Woerdman} (1994), Angular
  momentum and spin-orbit coupling for microwave photons, \textit{Optics
  Comm.}, \textit{104}, 229--233,
  \doi{doi}{http://dx.doi.org/10.1016/0030-4018(94)90547-9}.

\bibitem[{\textit{{LaBelle} and {Treumann}}(2002)}]{LaBelleTreumann:SSR:2002}
{LaBelle}, J., and R.~A. {Treumann} (2002), Auroral radio emissions, 1:
  {H}isses, roars, and bursts, \textit{Space Sci. Rev.}, \textit{101}(3),
  295--440.

\bibitem[{\textit{{Lavery} et~al.}(2013{\natexlab{a}})\textit{{Lavery},
  {Robertson}, {Sponselli}, {Courtial}, {Steinhoff}, {Tyler}, {Willner}, and
  {Padgett}}}]{LaveryA_2013_NJP}
{Lavery}, M.~P.~J., D.~J. {Robertson}, A.~{Sponselli}, J.~{Courtial}, N.~K.
  {Steinhoff}, G.~A. {Tyler}, A.~E. {Willner}, and M.~J. {Padgett}
  (2013{\natexlab{a}}), {Efficient measurement of an optical
  orbital-angular-momentum spectrum comprising more than 50 states},
  \textit{New Journal of Physics}, \textit{15}(1), 013024,
  \doi{doi}{http://dx.doi.org/10.1088/1367-2630/15/1/013024}.

\bibitem[{\textit{{Lavery} et~al.}(2013{\natexlab{b}})\textit{{Lavery},
  {Speirits}, {Barnett}, and {Padgett}}}]{LaveryA_2013_Science}
{Lavery}, M.~P.~J., F.~C. {Speirits}, S.~M. {Barnett}, and M.~J. {Padgett}
  (2013{\natexlab{b}}), Detection of a spinning object using light's orbital
  angular momentum, \textit{Science}, \textit{341}, 537--540,
  \doi{doi}{10.1126/science.1239936}.

\bibitem[{\textit{Leyser}(2001)}]{Leyser:SSR:2001}
Leyser, T.~B. (2001), Stimulated electromagnetic emissions by high-frequency
  electromagnetic pumping of the ionospheric plasma, \textit{Space Sci. Rev.},
  \textit{98}, 223--328.

\bibitem[{\textit{Leyser et~al.}(2009)\textit{Leyser, Norin, McCarrick,
  Pedersen, and Gustavsson}}]{Leyser:PRL:2009}
Leyser, T.~B., L.~Norin, M.~McCarrick, T.~R. Pedersen, and B.~Gustavsson
  (2009), Radio pumping of ionospheric plasma with orbital angular momentum,
  \textit{Phys. Rev. Lett.}, \textit{102}(6), 065004,
  \doi{doi}{http://dx.doi.org/10.1103/PhysRevLett.102.065004}.

\bibitem[{\textit{Li and Li}(2013)}]{LianlinFang_2013_PRE}
Li, L., and F.~Li (2013), Beating the {Rayleigh} limit:
  Orbital-angular-momentum-based super-resolution diffraction tomography,
  \textit{Phys. Rev. E}, \textit{88}, 033,205,
  \doi{doi}{http://dx.doi.org/10.1103/PhysRevE.88.033205}.

\bibitem[{\textit{Liska and Meinke}(1970)}]{LiskaMeinke:NZ:1970}
Liska, H., and H.~Meinke (1970), {Der experimentelle Nachweis der beiden
  elementaren Typen von Energiewirbeln in Wellenfeldern},
  \textit{Nachrichtentechnische Zeit.}, \textit{23}, 445--448.

\bibitem[{\textit{{Mair} et~al.}(2001)\textit{{Mair}, {Vaziri}, {Weihs}, and
  {Zeilinger}}}]{Mair&al:N:2001}
{Mair}, A., A.~{Vaziri}, G.~{Weihs}, and A.~{Zeilinger} (2001), Entanglement of
  the orbital angular momentum states of photons, \textit{Nature},
  \textit{412}, 313--316, \doi{doi}{http://dx.doi.org/10.1038/35085529}.

\bibitem[{\textit{McMorran et~al.}(2011)\textit{McMorran, Agrawal, Anderson,
  Herzing, Lezec, McClelland, and Unguris}}]{McMorran:Science:2011}
McMorran, B.~J., A.~Agrawal, I.~M. Anderson, A.~A. Herzing, H.~J. Lezec, J.~J.
  McClelland, and J.~Unguris (2011), Electron vortex beams with high quanta of
  orbital angular momentum, \textit{Science}, \textit{331}, 192--195,
  \doi{doi}{http://dx.doi.org/10.1126/science.1198804}.

\bibitem[{\textit{Mohammadi et~al.}(2010{\natexlab{a}})\textit{Mohammadi,
  Daldorf, Forozesh, Thid{\'e}, Bergman, Isham, Karlsson, and
  Carozzi}}]{Mohammadi&al:RS:2010}
Mohammadi, S.~M., L.~K.~S. Daldorf, K.~Forozesh, B.~Thid{\'e}, J.~E.~S.
  Bergman, B.~Isham, R.~Karlsson, and T.~D. Carozzi (2010{\natexlab{a}}),
  {Orbital angular momentum in radio: {M}easurement methods}, \textit{Radio
  Sci.}, \textit{45}, RS4007--1--RS4007--14,
  \doi{doi}{http://dx.doi.org/10.1029/2009RS004299}.

\bibitem[{\textit{Mohammadi et~al.}(2010{\natexlab{b}})\textit{Mohammadi,
  Daldorff, Bergman, Karlsson, Thid{\'e}, Forozesh, Carozzi, and
  Isham}}]{Mohammadi&al:IEEETAP:2010}
Mohammadi, S.~M., L.~K.~S. Daldorff, J.~E.~S. Bergman, R.~L. Karlsson,
  B.~Thid{\'e}, K.~Forozesh, T.~D. Carozzi, and B.~Isham (2010{\natexlab{b}}),
  {O}rbital angular momentum in radio: {A} system study, \textit{IEEE Trans.
  Antennas Propagat.}, \textit{58}, 565--572.

\bibitem[{\textit{Paterson}(2005)}]{Paterson:PRL:2005}
Paterson, C. (2005), Atmospheric turbulence and orbital angular momentum of
  single photons for optical communication, \textit{Phys. Rev. Lett.},
  \textit{94}(15), 153,901.

\bibitem[{\textit{{Poynting}}(1909)}]{Poynting:PRSL:1909}
{Poynting}, J.~H. (1909), The wave motion of a revolving shaft, and a
  suggestion as to the angular momentum in a beam of circularly polarised
  light, \textit{Proc.\ Royal Society London}, \textit{A82}, 560--567,
  \doi{doi}{http://dx.doi.org/10.1098/rspa.1909.0060}.

\bibitem[{\textit{{Rogers
  Corporation}}(2015)}]{RogersCorporation_2015_datasheet}
{Rogers Corporation} (2015), {RO4000} series high frequency circuit materials,
  \textit{Tech. rep.}, publication no.\ 92-004, url:
  https://\-www.\-rogerscorp.\-com/\-documents/\-726/\-acm/\-RO4000-Laminates---\-Data-sheet.\-pdf.

\bibitem[{\textit{Tamburini et~al.}(2012)\textit{Tamburini, Mari, Sponselli,
  Thid{\'e}, Bianchini, and Romanato}}]{TamburiniA_2012_NJP}
Tamburini, F., E.~Mari, A.~Sponselli, B.~Thid{\'e}, A.~Bianchini, and F.~Romanato
  (2012), Encoding many channels on the same frequency through radio vorticity:
  first experimental test, \textit{New Journal of Physics}, \textit{14}(3),
  033,001, \doi{doi}{http://dx.doi.org/10.1088/1367-2630/14/3/033001}.

\bibitem[{\textit{Thid{\'e} et~al.}(2007)\textit{Thid{\'e}, Then, Sj{\"o}holm,
  Palmer, Bergman, Carozzi, Istomin, Ibragimov, and
  Khamitova}}]{Thide&al:PRL:2007}
Thid{\'e}, B., H.~Then, J.~Sj{\"o}holm, K.~Palmer, J.~E.~S. Bergman, T.~D.
  Carozzi, Y.~N. Istomin, N.~H. Ibragimov, and R.~Khamitova (2007),
  Utlilization of photon orbital angular momentum in the low-frequency radio
  domain, \textit{Phys. Rev. Lett.}, \textit{99}(8), 087,701.

\bibitem[{\textit{Thid{\'e} et~al.}(2011)\textit{Thid{\'e}, Lindberg, Then, and
  Tamburini}}]{ThideA_2011_arXiv}
Thid{\'e}, B., J.~Lindberg, H.~Then, and F.~Tamburini (2011), Linear and
  angular momentum of electromagnetic fields generated by an arbitrary
  distribution of charge and current densities at rest, \textit{ArXiv
  e-prints}, p. 5 p.

\bibitem[{\textit{Turnbull et~al.}(1996)\textit{Turnbull, Robertson, Smith,
  Allen, and Padgett}}]{Turnball:OpticsComm:1996}
Turnbull, G.~A., D.~A. Robertson, G.~M. Smith, L.~Allen, and M.~J. Padgett
  (1996), The generation of free-space {L}aguerre-{G}aussian modes at
  millimetre-wave frequencies by use of a spiral phaseplate, \textit{Optics
  Comm.}, \textit{127}, 183--188.

\end{thebibliography}


% ...

\end{document}